\documentclass{llncs}
\usepackage[dvips]{graphicx}
\usepackage{amsmath}
\usepackage{amssymb}
\usepackage{verbatim}
\usepackage{textcomp}
\usepackage{amsfonts}

\author{M. V. Panduranga Rao\inst{1} \and V. Vinay \inst{2}}
\institute{Department of Computer Science and Automation \\Indian Institute of Science\\ Bangalore  560 012\\ India.\\
\email{pandurang@csa.iisc.ernet.in}
\and
PicoPeta Simputers Pvt Ltd\\
146, 5th Cross, RMV Extn\\
Bangalore - 560 080\\ India.\\
\email{vinay@picopeta.com}
\thanks{A preliminary version of this paper appeared in the proceedings of
workshop on Algorithms and Complexity in Durham (ACiD-2005).}
}

\date{}
\begin{document}
\newlength{\twidth}
\title{Quantum Finite Automata and Weighted Automata}
\maketitle
\begin{abstract}
Quantum finite automata derive their strength by exploiting interference in complex valued probability 
amplitudes.
Of particular interest is the 2-way model of Ambainis and
Watrous that has both quantum and 
classical states (2QCFA)  [A. Ambainis and J. Watrous, Two-way finite automata with quantum and classical state, 
Theoretical Computer Science, 287(1), pp. 299-311, 2002], since it combines the advantage of
the power of interference in a constant-sized quantum system with a 2-way head.

This paper is a step towards finding the least powerful model which is purely classical and can 
mimic the dynamics of quantum phase.
We consider weighted automata with the Cortes-Mohri definition of language recognition [C. Cortes and  M. Mohri, Context-Free 
Recognition with Weighted Automata, Grammars 3(2/3), pp. 133-150, 2000]  as a candidate model for simulating 2QCFA.

Given any 2QCFA that (i) uses the accept-reject-continue observable, (ii) recognizes a language with one-sided error and (iii) the entries of whose
unitary matrices are algebraic complex numbers, we show a method of constructing
a weighted automaton over $\mathbb{C}$ that simulates it efficiently.

\end{abstract}

\section{Introduction}
Quantum Finite Automata (QFA) have been an area of active research in  the recent past, with a lot models being investigated for their power. 
The first of such models, proposed by Moore and Crutchfield  \cite{MCr} and Kondacs and Watrous \cite{KW}, is the Measure-Once (MO) 1-way QFA.
In this model, the finite automaton has a read-only input tape
and a finite set of states $Q$ with $|Q|=k$, a constant. We associate a Hilbert space $\mathcal{H}$ with 
$Q=\{|q\rangle\}$ forming a basis.\footnote{See Section 2 for a quick introduction to quantum computing.} A state vector of the MO-1QFA is then a vector of unit norm in this space.
A series of unitary transformations, each depending on the input symbol, is applied on an initial quantum state 
as the input tape  is scanned from the beginning to the end. The final state vector is a linear superposition of the basis vectors:
$|\psi\rangle = \sum_{q\in Q}\alpha_q |q\rangle $, where $\alpha_q\in\mathbb{C}$ and $ \sum_{q\in Q}|\alpha_q|^2=1$.
On reaching the end of the input, a measurement is performed. If the state observed is among those designated as accepting, the input is accepted. 
Crutchfield and Moore \cite{MCr}  and Brodsky and Pippenger \cite{BP} showed that this model accepts only a proper subset of regular languages with
bounded error. 

A more powerful model, the Measure-Many (MM) 1-way QFA, was proposed by Kondacs and Watrous \cite{KW}. In this model, 
measurements are made after reading every input symbol. 
Further, $Q$ is partitioned into \emph{accepting, rejecting} and \emph{non-halting} subspaces. 
Measurements are carried out in such a way that the outcomes correspond to only the subspaces and not the individual states. 
If an accepting or rejecting subspace is observed, then the input is accepted or rejected respectively. 
If the outcome is 
\emph{non-halting}, the computation proceeds from a normalized vector in the non-halting subspace. 
Even this model has been shown to accept only a subset of regular languages in \cite{KW}. A lot of effort has gone into  characterizing the languages 
accepted by this model, cf. \cite{AK,AKV,BP}. 
A still more powerful model, the 2-way quantum finite automaton (2QFA) was also proposed by Kondacs and Watrous \cite{KW}. This model allows superpositions 
where the head can be in many positions simultaneously on the input (of length $n$). Not only can this model recognize all regular languages,
but also some non-regular ones like $L_{eq}=\{a^mb^m\mid m\in\mathbb{N}\}$ with bounded error. However, the $O(\log n)$ qubits required to store the position of the head in the 
``finite" control make this model costly, and goes against the spirit of finite automata. 

Seeking to harness the advantages of both quantum states and the ability of the head of the tape to move both ways, Ambainis and Watrous \cite{AW} proposed
a model intermediate between 1QFA and 2QFA. They showed that this model, called the 2-way finite automata with quantum and classical states (2QCFA), 
can recognize  $L_{eq}$ and $L_{pal}= \{x \in \{a,b\}^*\mid x=x^R \}$, where $x^R$ is the reverse of $x$, with bounded error in polynomial and exponential time respectively. 

The only resource available to 2QCFA is the power to create interference in 
probability amplitudes. In this paper we look for an existing classical model that
can do the same.

In this paper we focus on 2QCFA with the reasonable restriction that the 
entries of the unitary operators acting on the quantum component are drawn from 
the field of algebraic numbers. We compare this variant, which we denote by 
2QCFA$(\mathbb{A})$, with weighted automata.

A weighted automaton is essentially a DFA with the transitions labelled by weights
drawn from a semiring in addition to symbols from a finite alphabet.\footnote{See next section for some basics of weighted automata. See also~\cite{Eilenberg,Schut}
for theory and~\cite{MPR} for applications.}
Cortes and Mohri~\cite{CM} defined a notion of language
recognition by weighted automata based on the sum of path-weights labeled by the 
input string.\footnote{In general, power of language recognition depends on the semiring 
used. Cortes and Mohri investigated this power for several
semirings and showed weighted automata for recognizing several classes
of context free languages.}

The motivation for comparing weighted automata with 2QCFA is that addition of 
path-weights can naturally capture interference in weighted automata. 
However, there are potential hitches. 
To begin with, the measurement operator does not have a parallel in weighted automata.
Secondly, moduli of complex numbers can not be obtained in weighted automata. How 
does one calculate, say,  $\sum_i|\alpha_i|^2$? And finally, one has to reconcile the
one-sided bounded error notion of language recognition for 2QCFA with the Cortes-Mohri
definition of language recognition for weighted automata.

In this paper we show that these problems can be surmounted when simulating
2QCFA that accept with one-sided error and whose unitary matrices have only 
algebraic complex numbers as entries. In other words, given a 2QCFA$(\mathbb{A})$
that recognizes a language $L$,
we show a method of constructing a weighted automaton over the complex semiring
that efficiently recognizes $L$.

This paper is organized as follows. The next section provides some useful 
definitions and some basics of 2QCFA and weighted  automata.
Section~\ref{blocksec3} gives a simulation of 2QCFA$(\mathbb{A})$ by  weighted automata.

\section{Preliminaries}\label{blocksec2}

We begin this section with an introduction to quantum computing. For details, 
please see the text by Neilson and Chuang~\cite{NC}.
A superposition of $k$ states $\{q_0,q_1,\ldots, q_{k-1}\}$ is a vector
of unit norm in a $k$- dimensional Hilbert space $\mathcal{H}$, with 
$Q=\{|q_0\rangle,|q_1\rangle,\ldots, |q_{k-1}\rangle\}$ serving as a basis of 
elementary unit vectors. Thus, a linear superposition may be written as 
$|\psi\rangle = \sum_{i=0}^{k-1}\alpha_i|q_i\rangle$ with each $\alpha_i \in \mathbb{C}$ and $\sum_{i=0}^{k-1}|\alpha_i|^2=1$.
A unitary operator on $\mathcal{H}$ is a norm preserving linear operator. 
Application of a unitary operator $U$ on a superposition $|\psi\rangle$
evolves the system to $U|\psi\rangle$.
An orthogonal measurement of the system is specified by a set $\{P_i\}$
of operators such that (i) $P_i=P_i^\dagger$ for each $i$, where $P_i^\dagger$
denotes the adjoint of $P_i$ (ii) $P_i^2 =P_i$ for each $i$ (iii) $P_iP_j=0$
for $i\neq j$ and (iv) $\sum_i P_i =I$.
Measuring a superposition $|\psi\rangle$ through such a set yields $i$
with a probability $||P_i |\psi\rangle ||^2 $ for each $i$. The superposition
itself collapses to $\frac{1}{||P_i|\psi\rangle ||} P_i|\psi\rangle$ for the $i$
that was the yield.
The subspaces $E_i$ of $Q$ on which the projectors $P_i$ operate partition $Q$; 
such a partition forms an \emph{observable}.
\newpage
\subsection{Automata with Classical and Quantum States}
Formally, a 2QCFA $M$ is a $9$ tuple:
$M=(Q,S,\Sigma,\Theta,\delta,q_0,s_0, S_{acc},S_{rej})$
where
\begin{itemize}
\item $Q$ and $S$ are finite sets of quantum and classical states respectively.
\item $\Sigma$ is a finite alphabet, $\Gamma= \Sigma \bigcup \{$\textcent$,\$\}$ being the tape alphabet with \textcent,$\$ \notin \Sigma$ as left and right
end-markers respectively. 
\item $q_0$ and $s_0$ are the initial quantum and classical states respectively.
\item $S_{acc}$, $S_{rej} \subseteq S$ are the subsets of classical accepting and rejecting states respectively.
\item $\Theta$ is a two-parameter transition function which governs the evolution of the quantum state.
$\Theta(s,\sigma)$, where $s\in S\backslash(S_{acc}\bigcup S_{rej})$ and  $\sigma \in \Gamma$,
is either a unitary transformation or a measurement.
\item $\delta$ is a function describing the evolution of the classical state and is defined as follows:

Unless $\Theta(s,\sigma)$ is a measurement, $\delta(s,\sigma) \in S\times D$, for $D=\{-1,0,1\}$ where $-1,0,$ and $1$ specify the movement of the tape head.

If $\Theta(s,\sigma)$ is a measurement, $\delta$ also takes the outcome of the measurement
into account.
\end{itemize}

\subsection{Weighted Automata}
Weighted automata are essentially DFA with weights
(drawn, in general, from a semiring;\footnote{
A system $(\mathbb{K},\oplus,\otimes,\bar{0}, \bar{1})$ is a 
right semiring if:
\begin{itemize}
\item $(\mathbb{K},\oplus,\bar{0})$ is a commutative monoid with $\bar{0}$ as the
identity element for $\oplus$.
\item $(\mathbb{K},\otimes,\bar{1})$ is a monoid with $\bar{1}$ as the identity element for $\otimes$.
\item $\forall a,b,c\in \mathbb{K}$, $(a\oplus b)\otimes c= (a\otimes c)\oplus (b\otimes c)$, and
 $a\otimes(b\oplus c)=(a\otimes b) \oplus (a\otimes c)$
\item $\forall a\in\mathbb{K}$, $a\otimes\bar{0} = \bar{0}\otimes a = \bar{0}$.
\end{itemize}} in our case from $\mathbb{C}$)
attached to the transitions, in addition to the symbols from a finite alphabet
$\Sigma$. There is a set of initial and final states. A weighted automaton $W$
associates an input $x\in \Sigma^*$ to an element $W\circ x$ in $\mathbb{C}$
calculated as the sum of products of the weights along every path from the 
initial to the final states. Cortes and Mohri~\cite{CM} defined an elegant
notion of language recognition in this setting. If the element $W\circ x$ so associated
with the input falls in a predefined subset $\mathbb{J}$ of $\mathbb{C}$,
we say that the input is accepted by the weighted automaton $W$. Thus,
in addition to the actual evaluation of $W\circ x$, the time complexity 
of deciding whether $x$ belongs to the language of $W$ also depends on the complexity of
deciding membership in $\mathbb{J}$.
Under this definition of language recognition, nondeterministic finite automata 
and probabilistic finite automata can be seen as specific instances of general
weighted automata over different semirings.

While most applications use  1-way weighted automata, we define and use a 2-way version. The definition is similar, 
except that now the machine reads input off a tape and the head can travel in both directions.
A 2-way weighted automaton is formally defined as follows.

\begin{definition} A 2-way weighted finite automaton over $\mathbb{C}$ is a $7$-tuple

$W= (S,\Sigma,I,F,\Delta,\lambda,\rho)$
where
\begin{itemize}
\item $S$ is a finite set of states.
\item $\Sigma$ is a finite alphabet, $\Gamma= \Sigma \bigcup \{$\textcent$,\$\}$ being the tape alphabet.
\item $I,F \subseteq S$ are sets of initial and final states respectively.
\item $\Delta \subseteq S\times \Gamma\times \mathbb{C}\times S \times D $ is a finite set of transitions, where $D=\{-1,0,+1\}$ is the direction of
movement of the head on the tape.
\item $\lambda:I\rightarrow \mathbb{C}$ and $\rho :F \rightarrow \mathbb{C}$. 
 \end{itemize}
 \end{definition}

A weighted automaton recognizes languages in the following manner.
For a transition $e\in \Delta$ we denote by $w[e]$ its weight and by $i[e]$ its input label. 
We denote a \emph{path} $e_1\ldots e_k \in \Delta^*$ by $\pi$.
$first[\pi]$ is the originating state of the transition $e_1$ and $last[\pi]$ is the destination  state of $e_k$.
The weight of a path $\pi$ is given by $w[\pi] = w[e_1]\otimes \ldots \otimes w[e_k]$.
If $\Delta$ is such that the movement of the head is always deterministic,\footnote{Head movement is deterministic if the transitions are such that at every time
step the head moves in the same direction, no matter what state the machine is in.}
as will be the case in our case, then we can define
$x'= scan_W(x)$ as the sequence symbols of the input $x$ to come under the head.
It is important to note that since the head is 2-way, 
the same symbol of the input may occur more than once in $scan_W(x)$.
We denote by $P(q,q')$ the set of all paths between states $q$ and $q'$.
If $P(q,q')$ is the set of all paths between states $q$ and $q'$, we say
\[
	P(I,F) = \bigcup_{q\in I, q'\in F} P(q,q').
\]
Then we can also define
\[
\Pi(x') =\{\pi \in P(I,F) | label[\pi] = x'\},
\]
where $x'= scan_W(x)$ is the sequence symbols of the input $x$ to come under the head.
Moreover, for our purpose, a single initial state $q_0$ suffices: $|I|=1$.

The input is associated to a weight in $\mathbb{C}$ by the automaton $W$ as
\begin{equation}
W\circ x' = \sum_{\pi \in\Pi(x')} \lambda(first[\pi])\otimes w[\pi] \otimes \rho(last[\pi])
\end{equation}

With this in hand, we can now define language recognition by a weighted automaton.

\begin{definition}
Let $\mathbb{J} \subseteq \mathbb{C}$. We say that a string $x\in \Sigma^*$ is $\mathbb{J}$-recognized by
the weighted automaton $W$ if $W\circ scan_W(x)\in\mathbb{J}$.
\end{definition}

\section{QCFA and Weighted Automata}\label{blocksec3}
We begin with some definitions in the context of 2QCFA$(\mathbb{A})$ that will be useful in what follows.

\begin{definition}
If $\Theta(s,\sigma)$ is a unitary transformation for all 
$\sigma \in \Sigma$, we call $s\in S$ a \emph{unitary state}. We denote the subset of
such states by $S_u$. On the other hand, if $\Theta(s,\sigma)$ is a measurement
operation for any $\sigma \in \Sigma$, we call $s\in S$ a \emph{measurement} state and
denote the subset of such states by $S_m$.
\end{definition}

For the quantum part, we will use an obervable such that the outcome of a
measurement will tell if the computation halts immediately by accepting
or rejecting the input, or if it is to be continued.

To that end, we partition $Q$ into $Q_{acc}$, $Q_{rej}$ and $Q_{nh}$,
(for acceptance, rejection and continuation of computation respectively)
such
that $Q_{acc}\cap Q_{rej} =\phi$ and $q_0\in Q_{nh}$.
If $E_{acc}$, $E_{rej}$ and $E_{nh}$ are the subspaces spanned by states in 
$Q_{acc}$, $Q_{rej}$ and $Q_{nh}$ respectively, we use the observable defined
by $E_{acc}\oplus E_{rej} \oplus E_{nh}$. 
If the outcome of a measurement corresponds to $Q_{nh}$, computation continues
from a normalized vector in the subspace spanned by the states of $Q_{nh}$.
The observable for a ``final" measurement 
will not have a non-halting subspace.

We use the one-sided error notion of language acceptance for the 2QCFA$(\mathbb{A})$:
\begin{definition}
A 2QCFA$(\mathbb{A})$ $M$ is said to recognize a language $L$ with bounded one-sided error $\epsilon>0$
if 
\begin{eqnarray}
Prob[\textrm{ M accepts x}] &=& 1 \indent\indent \forall x \in L \nonumber \\
                           &\leq& \epsilon \indent\indent \forall x \notin L \nonumber \\ \nonumber
\end{eqnarray}

\end{definition}

We now state the main result of this paper. 
\begin{theorem}
Given a 2QCFA$(\mathbb{A})$ $M$ that recognizes a language $L$ with one-sided 
error $\epsilon>0$, there exists a 2-way weighted automaton $W$ 
that accepts $L$ in time $O(| scan_W{x}|)$.
\end{theorem}
\begin{proof}
Given a 2QCFA$(\mathbb{A})$ $M_A=(Q_A,S_A,\Sigma,\Theta,\delta_A,q_0,s_0,S_{acc},S_{rej})$ that
accepts a language $L$ with bounded error probability $\epsilon >0$,
we construct the weighted automaton $W=(S_W,\Sigma,q^0_0,F,\Delta,1,1)$ that also accepts $L$. The set $F$ of final states will have the cardinality of
$|S_m||Q_{rej}|$, as we will see soon.
We will denote an edge in $\Delta$ of a weighted automaton by the tuple $(s_1,\sigma,w,s_2,d)$: 
in the state $s_1$, on seeing the symbol $\sigma$, the finite control moves to
the state $s_2$, the head by $d\in\{-1,0,+1\}$, and the weight of this
transition is $w$.

\subsection*{The Construction}

\begin{enumerate}
\item The set $S_W$ is the union of the disjoint sets $Q_P$, $S_P$ and $F$ 
defined as: 

(a) $Q_P= \{q^i_{j_1}\}$ for $0\leq i\leq |S_A|-1$ and $0\leq j_1\leq |Q_A|-1$.
Thus we have $|S_A|$ copies of $Q_{acc}$, $Q_{rej}$ and $Q_{nh}$, indexed by $i$
as $Q^i_{acc}$, $Q^i_{rej}$ and $Q^i_{nh}$ respectively.

(b) $S_P= \{s^i_{j_2}\}$ for $0\leq i\leq |S_A|-1$ and $0\leq j_2\leq |Q_A|-1$.

(c) $F=\{f^i_j\}$ for $i$ such that $s_i\in S_m$ and $j$ such that
$q^i_j \in Q^i_{rej}$.

\item If $s_i\in S_u$ and $\Theta(s_i,\sigma)$ is a unitary
transformation  of the form:
\[
U:|q_{j_1}\rangle \rightarrow \sum_{j_2=0}^{|Q|-1} \alpha^{j_1}_{j_2} |q_{j_2}\rangle
\]
for all $q_{j_1} \in Q$, then 
add the edge $(q^i_{j_1},\sigma,\alpha^{j_1}_{j_2},s^i_{j_2},0)$
for $0\leq j_1,j_2 \leq |Q|-1$.

\item If $s_i \in S_m$ and $\Theta(s_i,\sigma)$ is a unitary transformation,
we proceed as in the case of $S_u$. For $s_i \in S_m$ and $\sigma \in \Sigma$
such that $\Theta(s_i,\sigma)$ is a  measurement, add the following edges:
\begin{itemize}
\item $(q^i_j,\sigma,0,q^i_j,0)$ for all $q^i_j\in Q^i_{acc}$,
\item $(q^i_j,\sigma,1,s^i_j,0)$ for all $q^i_j\in Q^i_{nh}$, and
\item $(q^i_j,\sigma,e^{\rho^i_j},f^i_j,0)$ for all $q^i_j\in Q^i_{rej}$, where
$f^i_j \in F$, $e$ is the base of the natural logarithm, and $\rho^i_j$ are $|S_m||Q_{rej}|$ distinct algebraic numbers.
\end{itemize}
\item If $\delta_A(s_i,\sigma) = (s_{i'},d)$, for $s_{i'}\in S_A$,
then add the edge $(s^i_{j_1},\sigma,1,q^{i'}_{j_1},d)$.

\end{enumerate}

This completes the construction. For the sake of brevity and clarity, we did not
spell out the construction for a ``final" measurement state. 
It is similar, except that now there is no $Q^i_{nh}$.  

{\bf Some important observations:}
\begin{enumerate}
\item The head does not move during transitions from states in
$\{q^i_{j_1}\}^{|Q_A| -1}_{j_1=0}$ to states in $\{s^i_{j_2}\}^{|Q_A| -1}_{j_2=0}$.

\item The states in the set $\{q^i_{j_1}\}^{|Q_A| -1}_{j_1=0} \cup \{s^i_{j_2}\}^{|Q_A| -1}_{j_2=0}$ 
may be seen as belonging to the same \emph{block} labelled $i$. Since $S_A$ is a 
set of deterministic states, by step 4 of the construction, so is $S_P$.
Thus, at any point in time, the machine $W$ is in the states of the same block.

\item Intra-block transitions are derived from $\Theta$, while 
inter-block transitions are derived from $\delta_A$.

\item The number of states in the blocks
is actually an overkill, since there are also classical states in 
the 2QCFA$(\mathbb{A})$ that do not play a role in the evolution of the quantum part.
By our construction, the blocks corresponding to such states will have trivial
internal transitions: $(q^i_{j_1},1,\sigma,s^i_{j_1},0)$. 

\end{enumerate}

Suppose $|\psi \rangle = \sum_{j=0}^{k-1}\gamma_j(x')|q_j \rangle$ is the state vector of the 2QCFA$(\mathbb{A})$ after reading the input $x'=scan_{M_A}(x)$.
Consider a sequence of unitary transforms uninterrupted by measurements. The
block structure then essentially simulates (unitary) matrix multiplication. If 
$i$ is the latest block entered, it is easy to see that $\gamma_j(x') = \sum_{\pi \in \Pi^i_j(x')}w[\pi]$, where
\[
\Pi^i_j(x') =\{\pi \in P(q^0_0,s^i_j) | label[\pi] = x'\}.
\]

Consider now a computation wherein measurements have occurred without resulting
in termination till some point in time. The fact that it has not terminated 
implies that at every measurement the quantum part collapsed to a vector
in the subspace spanned by $Q_{nh}$. Recall that if $|\psi\rangle$ is the 
state vector before measurement, it collapses to 
$\frac{P_{nh}|\psi\rangle}{\sqrt{\langle\psi|P_{nh}|\psi\rangle}}$ post 
measurement, where $P_{nh}$ is the projection operator onto $Q_{nh}$.
While step 3 in the construction ensures that the relative amplitudes
and phases of the 2QCFA$(\mathbb{A})$ are preserved in the simulating weighted automaton,
it is not possible to mimic division by the overall normalization factor 
$\sqrt{\langle\psi|P_{nh}|\psi\rangle}$. These normalization factors can be accumulated over several measurements and clubbed together.
This leads to the following lemma.

\begin{lemma}
Let $| \psi \rangle = \sum_{j=0}^{k-1}\gamma_j(x')| q_j \rangle$ be the state vector of the 2QCFA$(\mathbb{A})$ after reading the input $x'=scan_{M_A}(x)$. Then, 
\[
\gamma_j(x') = 1/\mathcal{P}\sum_{\pi \in \Pi^i_j(x')}w[\pi] 
\]
where $\Pi^i_j(x')=\{\pi \in P(q^0_0,s^i_j) | label[\pi] = x'\}$  and $\mathcal{P}$ is an overall normalization
factor.
\end{lemma}

Since the normalization factor is common to all probability amplitudes, 
it does not pose a problem as far as interference is concerned.

\begin{lemma}
Let $\Pi(x')$ be the set $\{\pi\in P(q_0^0,F)| label[\pi]=x'\}$.
Then,
\[
A\circ x' = \sum_{\pi\in \Pi(x')}w(\pi)=0
\]
if and only if $x\in L$.
\end{lemma}
\begin{proof}
If $x\in L$, the 2QCFA$(\mathbb{A})$ accepts $x$ with certainty: the probability amplitude for
observing any state in $Q_{rej}$ is zero. By the construction and
the previous lemma, 
$w^i_j=0$ for all $q^i_j\in Q^i_{rej}$. Therefore the sum $\sum_jw^i_je^{\rho^i_j}$ of 
weights of all paths from $q^0_0$ to the final states $f^i_j$ 
will be zero for all $i$ such that $s_i\in S_m$. 

For the other direction, we begin by noting a classic result of Lindemann (see~\cite{Niven}):
\begin{theorem}
Given any distinct algebraic numbers $\phi_1, \phi_2,\ldots,\phi_n$, the values $e^{\phi_1}, e^{\phi_2}, \ldots,e^{\phi_n}$ are linearly
independent over the field of algebraic numbers.
\end{theorem}

If $x\notin L$, there exists a non-zero amplitude for some state in $Q_{rej}$.
Again by the above lemma, this means that there will be non-zero
weights $w^i_j$ on some paths to states in $Q^i_{rej}$, for some $i$ such that 
$s_i\in S_m$. 
Consider any such $i$. The sum of weights of all paths from 
$q^0_0$ to the final states $f^i_j$ in this block is $\sum_jw^i_je^{\rho^i_j}$,
not all $w^i_j$ being  zero.
Then, by Lindemann's theorem, this sum is not zero.

Thus, if $w^i_j$ are the weights of the paths ending in the rejecting quantum
states $q^i_j\in Q^i_{rej}$ of a block corresponding to a measurement state,
\[
\sum_jw^i_je^{\rho^i_j} = 0 \textrm{ \quad if and only if $w^i_j=0$ for all $w^i_j$,}
\]
and the lemma follows. $\square$
\end{proof}

Therefore, checking whether an input $x$ is in $L$ amounts to checking if $A\circ x'$ equals $0$.
If it does, we accept the string, else we reject it. Hence, the constructed automaton $0$-recognizes $L$. It is significant that the $\mathbb{J}$ is such a small subset of 
$\mathbb{J}$, as it reduces the membership test to constant time. The blow-up in the size of the machine is also only by a constant factor. So, the total
time taken is $O(|x|)$, owing to a constant number of scans. Hence the theorem.
$\square$
\end{proof}

\subsection{Examples}
We illustrate the construction for  the 2QCFA$(\mathbb{A})$ of Ambainis and Watrous~\cite{AW} that recognizes
palindromes. For convenience, a three dimensional quantum part with real amplitudes is used in \cite{AW}. Initially, the quantum part is in the state $| q_0 \rangle$.
In the first scan, on reading an ``a", the matrix $A$ is applied and
$B$ on a ``b". In the second scan, $A^{-1}$ on reading an ``a" and $B^{-1}$ on reading a ``b" where 
\begin{displaymath}
\mathbf{A} = 
\frac{1}{5}\left( \begin{array}{ccc}
4 & 3 & 0 \\
-3 & 4 & 0 \\
0 & 0 &  5 \\
\end{array} \right)
\mathbf{B} = 
\frac{1}{5}\left( \begin{array}{ccc}
4 &  0 & 3 \\
0  & 5 & 0 \\
-3 & 0 &  4 \\
\end{array} \right).
\end{displaymath}
After completion of the two scans, a measurement is performed.
If the input is a palindrome, then $q_0$ will be observed every time. Otherwise,
either $q_1$ or $q_2$ will be observed with a small probability.

Two classical states are required for the two scans. Hence, the corresponding weighted automaton has two ``non-trivial" blocks, one for each scan
by the 2QCFA. The weights in one block are taken from $A$ and $B$ and in the other, from
$A^{-1}$ and $B^{-1}$.

Figure 1 shows the state transition diagram of the weighted automaton
simulating the subroutine.
In each box, the states to the left belong to $Q_P$ and the transitions
going out of them are labelled with probability amplitudes. The states
to the right belong to $S_P$. Since the classical transitions are 
deterministic in the 2QCFA$(\mathbb{A})$ being simulated, so are the ones going out of these
states, by construction. Note that a ``trivial block",
the block corresponding to the classical state of the machine between the two passes,
in which the head is restored to the left end-marker, has not been shown in the 
figure.

\begin{figure}[!ht]\label{fig1}
\begin{center}
\includegraphics[width=0.9\textwidth]{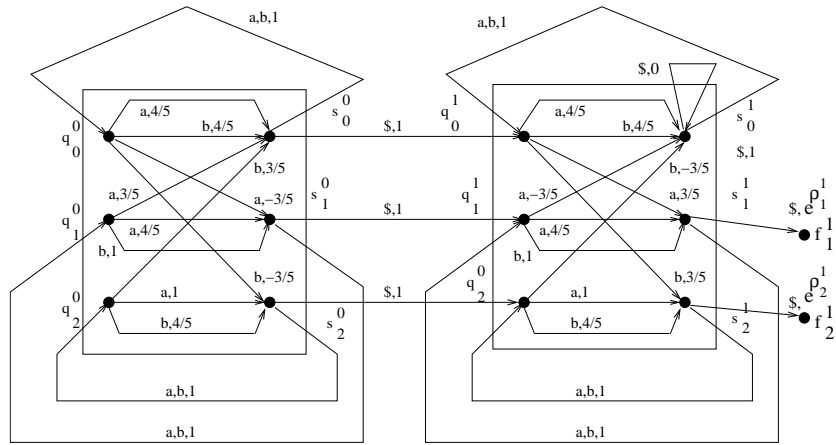}
\caption{Recognizing palindromes.}
\end{center}
\end{figure}

\bibliographystyle{plain}
\bibliography{../../thesis/7_11/bib}

\end{document}